\documentclass[aps,pra,groupedaddress]{revtex4}

\usepackage{graphicx}
\begin{document}


\title{
Violation of local uncertainty relations
as a signature of entanglement}

\author{Holger F. Hofmann}
\email{h.hofmann@osa.org}
\author{Shigeki Takeuchi}
\affiliation{PRESTO, 
Japan Science and Technology Corporation 
(JST)\\
Research Institute for Electronic Science, Hokkaido 
University\\
Kita-12 Nishi-6, Kita-ku, Sapporo 060-0812, Japan}

\date{\today}

\begin{abstract}
Entangled states represent correlations 
between two separate systems that are too precise 
to be represented by products of local quantum states.  
We show that this limit of precision for the local
quantum states of a pair of N-level systems can be 
defined by an appropriate class of uncertainty 
relations. The violation of such local 
uncertainty relations may be used as an experimental
test of entanglement generation.
\end{abstract}


\maketitle

\section{Introduction}
As more and more experimental realizations of entanglement
sources become available, it is necessary to develop 
efficient methods of testing the entanglement produced by 
such sources \cite{Mai01,Vaz02,Lam01,How02,Hag97,Guh02}. 
In particular, the output of entanglement sources is 
usually in a mixed state due to various decoherence effects. 
For such mixed states, it can be a 
difficult task to distinguish whether the output is really
entangled, or whether it is separable into some mixture
of non-orthogonal product states. Although there are simple 
formal criteria if the complete density matrix is known, 
the experimental determination of all matrix elements 
of an output state requires considerable experimental
efforts \cite{Whi99}. It is therefore desirable to simplify
the verification of entanglement by reducing it to the observation
of only a few characteristic statistical properties. One well 
known statistical property of entanglement is the violation 
of Bell's inequalities, and previous experiments often relied 
on this property as proof of entanglement \cite{How02,Wei98}. 
However, the requirements for Bell's inequality violations are 
usually more restrictive than the conditions for entanglement 
\cite{Wer89}, and the experiments still require a comparison of
at least four different correlation measurements. 
These complications arise from the fact that Bell's inequalities 
test the possibility of local hidden variable models. 
For entanglement verification, it is not necessary to exclude
hidden variable models, since entanglement can be defined 
entirely within the context of conventional quantum theory,
without any reverence to alternative models.
A more efficient method may therefore be the definition of a 
boundary between entangled states and non-entangled states 
in terms of expectation values of special operators called 
entanglement witnesses \cite{Hor96,Lew00}. 
Each witness operator defines a statistical limitation for
separable states derived directly from the topology of
Hilbert space.
However, the construction of witnesses that can be tested 
with only a few local von Neumann measurements is still a 
highly non-trivial task \cite{Guh02}. Since the experimental
verification of optical entanglement typically uses local 
von Neumann measurements, it may be desirable to express the 
criteria for entanglement directly in terms of the measurement
statistics obtained in such experiments.

In this paper, we therefore propose an alternative approach to 
entanglement verification based on the observation that 
entanglement seems to overcome the uncertainty limit by 
allowing correlations between sets of non-commuting properties 
of two systems to be more precise than any local definition 
of these properties could ever be. 
Since this precision in the 
correlations between two spatially separated systems is the 
property that originally lead to the discovery and definition 
of entanglement \cite{EPR,Cat}, a quantitative evaluation of
local uncertainty violations may provide one of the most
precise experimental measures of entanglement. 
A generalized characterization of entanglement as a 
suppression of noise below the local quantum limit may also be 
useful in the study of teleportation errors and related problems
of quantum communication \cite{Pop94,Hof02a,Tsu02}
and in the evaluation of the increased precision achieved by
applications of entanglement such as 
quantum lithography \cite{Bot00} or atomic clock 
synchronization \cite{Chu00,Joz00}. 

In the following, we first reformulate the uncertainty principle,
adapting it to arbitrary properties of N-level systems.
This reformulation of uncertainty provides unconditional 
limitations for the predictabilities of measurement outcomes
for any selection of non-commuting physical properties.
We can then derive local uncertainty limits valid for all 
non-entangled states. Since no separable
quantum state can overcome these limits, any violation 
of such local uncertainty relations is an unambiguous proof 
of entanglement. Some typical examples are provided and the
possibility of obtaining a quantitative measure of entanglement
from local uncertainties is discussed.

\section{Sum uncertainty relations for N-level systems}
\label{sec:ur}
The use of uncertainty arguments to study entanglement is 
well known from continuous variable systems 
\cite{EPR,Rei89,Dua00}. 
However, these arguments are based on the conventional 
product uncertainty of position and momentum.
This product uncertainty is based on the observation that
an eigenstate of position must have infinite momentum 
uncertainty and vice versa. In N-level systems, all physical
properties have upper and lower bounds, making infinite
uncertainties impossible. Consequently, the products of 
N-level uncertainties will always be zero if the system is 
in an eigenstate of one of the properties concerned.
This means that the product uncertainties derived from the
commutation relations of operators do not provide a generally 
valid uncertainty limit for N-level systems. In order to 
obtain a quantitative definition of uncertainty limits, 
it is therefore necessary to reformulate and adapt the 
uncertainty principle to N-level systems.

%
%
In its most general form, the uncertainty principle states
that it is never possible to simultaneously predict the 
measurement outcomes for all observables of the system. 
In terms of quantum theory, the relevant observables of
the system are represented by a set of hermitian operators 
$\{\hat{A}_i\}$. The uncertainty of $\hat{A}_i$ for any
given quantum state is then defined as the statistical 
variance of the randomly fluctuating measurement outcomes,
\begin{equation}
\delta A_i^2 = \langle \hat{A}_i^2 \rangle - 
\langle \hat{A}_i \rangle^2.
\end{equation}
This positive property of the quantum state can only be
zero if the quantum state is an eigenstate of $\hat{A}_i$,
representing perfect predictability of the measurement outcome.
We can therefore conclude that a quantum state with zero
uncertainty in all the properties $\hat{A}_i$ must be a 
simultaneous eigenstate of all the operators in $\{\hat{A}_i\}$.
If there is no such simultaneous eigenstate, there must 
be a non-trivial lower limit $U>0$ for the sum of 
the uncertainties,
\begin{equation}
\label{eq:basic}
\sum_i \delta A_i^2 \geq U.
\end{equation}
The limit $U$ is defined as the absolute minimum 
of the uncertainty sum for any quantum state. It therefore
represents a universally valid limitation of the measurement
statistics of quantum systems. 

Since $U$ represents a global
minimum, it may be difficult to determine its value in cases
where the operators $\hat{A}_i$ have a complicated form. 
However, there are a number of significant cases where this
limit is fairly easy to determine. For N-level systems, one
such fundamental limit can be obtained using the spin algebra of
the corresponding spin $l=(N-1)/2$ system with
\begin{eqnarray}
\label{eq:L^2}
(\hat{L}_x^2 + \hat{L}_y^2 + \hat{L}_z^2) \mid \psi \rangle 
= l(l+1) \mid \psi \rangle
\end{eqnarray} 
for any state $\mid \psi \rangle$. The expectation values
of $\hat{L}_i$ define a vector with a maximal length equal 
to the extremal eigenvalues of $\pm l$ along any axis.
We therefore obtain the uncertainty limit
\begin{equation}
\label{eq:Lrel}
\delta L_x^2 + \delta L_y^2 + \delta L_z^2 
= \underbrace{\langle \hat{L}_x^2 + 
\hat{L}_y^2 + \hat{L}_z^2 \rangle}_{= l(l+1)}
- \underbrace{\langle\hat{L}_x \rangle^2 + 
\langle\hat{L}_y \rangle^2 + 
\langle\hat{L}_z \rangle^2}_{\leq l^2}
\geq l.
\end{equation}
This uncertainty relation defines an absolute limit to the
precision of spin variables in any N-level systems. 
For the commonly studied case of two level systems,
the spin variables are often expressed in terms of 
the normalized Pauli matrices $\hat{\sigma}_i$, which 
have eigenvalues of $\pm 1$ instead of $\pm 1/2$. 
The sum uncertainty relation for the Pauli matrices
is then given by
\begin{equation}
\label{eq:sigrel}
\delta \sigma_1^2 + \delta \sigma_2^2 + \delta \sigma_3^2
\geq 2. 
\end{equation}
This uncertainty relation provides a quantitative description of
the fact that only a single spin component of a two level 
system can have a well defined value. It is also possible to 
formulate an uncertainty relation for only two spin
components by noting that $\delta \sigma_i^2 \leq 1$. This
simplified uncertainty relation reads
\begin{equation}
\label{eq:xyrel}
\delta \sigma_1^2 + \delta \sigma_2^2
\geq 1. 
\end{equation}
This is indeed the correct uncertainty minimum. For general
spin $l$ systems, such a simple derivation of the 
$\hat{L}_x$-$\hat{L}_y$ uncertainty is not possible, since the
maximal uncertainty of $\hat{L}_z$ is equal to $l^2$ and 
therefore exceeds the uncertainty limit for all three spin 
components. Nevertheless there exists an uncertainty limit
of $\hat{L}_x$ and $\hat{L}_y$ for any value of $l$, since 
$\hat{L}_x$ and $\hat{L}_y$ do not have any common 
eigenstates. For $l=1$, we have determined 
this limit by optimizing the spin squeezing properties of 
states with average spins in the $xy$-plane. The result reads
\begin{equation}
\label{eq:Lxyrel}
\delta L_x^2 + \delta L_y^2 \geq \frac{7}{16}.
\end{equation}
In the $\hat{L}_z$ basis, the minimum uncertainty state 
of this relation is given by
\begin{equation}
\mid \phi \rangle =
\frac{\sqrt{5}}{4} e^{-i \phi} \mid -1 \rangle
+  \frac{\sqrt{6}}{4} \mid 0 \rangle + 
\frac{\sqrt{5}}{4} e^{+i \phi} \mid +1 \rangle.
\end{equation}
It may be interesting to note that this minimum uncertainty
state has an $\hat{L}_z$-uncertainty 
of $\delta L_z^2=5/8$, so that the total of all three
spin uncertainties exceeds the limit set by relation 
(\ref{eq:Lrel}) by $1/16$. Relation (\ref{eq:Lxyrel})
is therefore more than just a truncated version of 
(\ref{eq:Lrel}).
\section{Local uncertainty limits}
\label{sec:lur}
It is now possible to apply the sum uncertainty relations to 
define the correlation limit for separable states. In general,
a pair of quantum systems A and B can be characterized 
by the operator properties $\hat{A}_i$ and $\hat{B}_i$
with the sum uncertainty relations given by
\begin{eqnarray}
\label{eq:ABlimit}
\sum_i \delta A_i^2 &\geq& U_A
\nonumber \\
\sum_i \delta B_i^2 &\geq& U_B.
\end{eqnarray}
It may be worth noting that the two Hilbert spaces of
system A and system B do not need to have the same 
dimension. In principle, local uncertainty limits can be
derived for any NxM system. Nor is it necessary that there
exists any specific relation between the properties
$\hat{A}_i$ and $\hat{B}_i$ other than that there is exactly
one property $\hat{A}_i$ in A for every property 
$\hat{B}_i$ in B.
The operator properties $\hat{A}_i+\hat{B}_i$ then
define a set of joint properties of the two systems that can
determined by local measurements of $\hat{A}_i$ and
$\hat{B}_i$, respectively. 
For product states, the measurement
values are uncorrelated and the uncertainties of 
$\hat{A}_i+\hat{B}_i$ are equal to the sum of the local
uncertainties,
\begin{eqnarray}
\delta (A_i+B_i)^2 &=& \delta A_i^2 + \delta B_i^2 
\nonumber \\
\mbox{for} \hspace{0.5cm} \hat{\rho} &=& 
\rho(\mbox{A}) \otimes \rho(\mbox{B}).
\end{eqnarray}
Therefore, the measurement statistics of product states
are limited by the uncertainty relation
\begin{equation}
\label{eq:plimit}
\sum_i \delta (A_i+B_i)^2 \geq U_A + U_B.
\end{equation}
Moreover, this uncertainty limit also applies to all
mixtures of product states, since the uncertainties of 
a mixture are always equal to or greater than the 
averaged uncertainties of the components. For 
the general case of $\hat{\rho}=\sum_m p_m \hat{\rho}_m$ and an 
arbitrary property $\hat{S}$, this relation between the 
uncertainties of a mixture and the uncertainty of its 
components can be obtained from
\begin{eqnarray}
\label{eq:sumrule}
\delta S^2 &=& \sum_m p_m \mbox{tr}\left\{
\rho_m (\hat{S}-\langle \hat{S} \rangle)^2
\right\}
\nonumber \\
&=& \sum_m p_m \left(
\underbrace{(\mbox{tr}\{\rho_m \hat{S}^2\}-
\mbox{tr}\{\rho_m \hat{S}\}^2)}_{= \delta S_m^2}
+\underbrace{(\mbox{tr}\{\rho_m \hat{S}\}
-\langle \hat{S} \rangle)^2}_{\geq 0}
 \right) \geq \sum_m p_m \delta S_m^2.
\end{eqnarray}
It follows from this result that the uncertainty relation
(\ref{eq:plimit}) for product states also applies to 
a mixture of product states,
\begin{eqnarray}
\label{eq:lur}
\sum_i \delta (A_i+B_i)^2 &\geq& U_A + U_B 
\nonumber \\
\mbox{for any} \hspace{0.2cm} \hat{\rho} &=& 
\sum_m p_m \; \rho_m(\mbox{A}) \otimes \rho_m(\mbox{B}).
\end{eqnarray}
Any violation of this uncertainty limit therefore proofs
that the quantum state cannot be separated into
a mixture of product states. However, entangled states 
can overcome this limitation, since entanglement
describes correlations that are more precise than the ones 
represented by mixtures of product states.
The violation of any local uncertainty relation of the
form (\ref{eq:lur}) is therefore a sufficient condition 
for the existence of entanglement.

\section{Violation of local uncertainty relations}
To illustrate how entanglement can overcome the 
local uncertainty limit defined by (\ref{eq:lur}),
it may be useful to consider the properties of 
maximally entangled states. Using the Schmidt bases
$\mid n \rangle_{\mbox{A}}$ and 
$\mid n \rangle_{\mbox{B}}$ for
A and B, these states can be written as
\begin{equation}
\mid E_{\mbox{max.}} \rangle_{\mbox{A;B}} = 
\frac{1}{\sqrt{N}}
\sum_n \mid n; n \rangle_{\mbox{A;B}}. 
\end{equation}
Such maximally entangled states appear to violate the 
uncertainty principle because any property of system A 
can be determined by a corresponding measurement on 
system B. 
That is, a measurement of an eigenvalue of $\hat{A}_i$
in A projects the quantum state in B into the eigenstate
of $-\hat{B}_i$ with the same eigenvalue as the one obtained
for $\hat{A}_i$ in A.
This means that, for any set of operators $\hat{A}_i$ 
in A, there is a set of corresponding operators 
$-\hat{B}_i$ in B such that the measurement result of 
$\hat{A}_i$ is always equal to the measurement result 
of $-\hat{B}_i$. In more formal terms, 
$\mid E_{\mbox{max.}} \rangle_{\mbox{A;B}}$ is a
simultaneous eigenstate of all $\hat{A}_i+\hat{B}_i$ with 
eigenvalues of zero \cite{Hof02b}.
Maximally entangled states can thus have a total 
uncertainty of zero in all properties $\hat{A}_i+\hat{B}_i$, 
maximally violating the uncertainty relation (\ref{eq:lur})
with
\begin{eqnarray}
(\hat{A}_i+\hat{B}_i)
\mid E_{\mbox{max.}} \rangle_{\mbox{A;B}} = 0 \hspace{0.3cm} 
&\mbox{and}& \hspace{0.3cm} 
\sum_i \delta (A_i+B_i)^2 = 0
\nonumber \\
\mbox{for} \hspace{0.3cm} 
\langle n \mid \hat{B}_i \mid n^\prime \rangle \! &=& \!
-\langle n^\prime \mid \hat{A}_i \mid n \rangle.
\end{eqnarray}
Experimentally, 
it is then possible to evaluate how close a given mixed state
output is to an intended maximally entangled state by measuring 
the remaining uncertainty due to imperfections in the entanglement
generation process. To obtain a quantitative estimate of the 
quality of entanglement generation, the measured uncertainty can 
be compared with the uncertainty limit of $U_A+U_B = 2 U$ for 
separable states. Specifically, the relative violation of local
uncertainty may be defined as
\begin{equation}
\label{eq:level}
C_{\mbox{\small LUR}}=1-\frac{\sum_i \delta (A_i+B_i)^2}{2 U}. 
\end{equation}
Since some amount of entanglement is necessary to overcome
the uncertainty limit, $C_{\mbox{\small LUR}}$ provides a 
quantitative
estimate of the amount of entanglement verified by the
violation of local uncertainty.  
In particular, it may be interesting to determine the minimal
amount of entanglement necessary to obtain a given value of
local uncertainty violation $C_{\mbox{\small LUR}}$ for 
various local uncertainty relations. Once such relations are
known, it will be possible to obtain reliable estimates
of entanglement from local uncertainty violations without 
additional assumptions about the quantum state.
\section{Application to entanglement between two spin-1 systems}
In general, any uncertainty relation of the type given by
(\ref{eq:basic}) can be used to define an uncertainty
limit for non-entangled states according to relation
(\ref{eq:lur}). However, in most cases it will be convenient
to define the limit in a highly symmetric way. This can
be achieved for any N-level system by using the spin 
uncertainty (\ref{eq:Lrel}). The local uncertainty relation 
for separable states of two spin $l=(N+1)/2$ systems is
given by
\begin{equation}
\label{eq:lcond}
\delta(L_x(\mbox{A})+L_x(\mbox{B}))^2+
\delta(L_y(\mbox{A})+L_y(\mbox{B}))^2+
\delta(L_z(\mbox{A})+L_z(\mbox{B}))^2 \geq 2 l.
\end{equation}
Any state that violates this uncertainty relation must be 
entangled. The optimal result of zero total uncertainty is 
obtained for the singlet state, defined by 
\begin{equation}
(\hat{L}_i(\mbox{A}) + \hat{L}_i(\mbox{B}))\mid \mbox{singlet} 
\rangle_{\mbox{A;B}} = 0.
\end{equation}
Experimental methods of generating such singlet
states for three level systems ($l=1$) have been 
realized using optical parametric 
downconversion to create photons entangled in their
spatial degrees of freedom \cite{Mai01,Vaz02}, or to create
entanglement between the polarization properties of 
a pair of two photon states \cite{Lam01,How02}. The relative
violation of local uncertainties defined by equation
(\ref{eq:level}) may serve as an easily 
accessible quantitative measure of the achievements 
represented by these experiments. 

In order to minimize the experimental effort involved in
characterizing the entanglement of three level systems, it
is also possible to use the local uncertainty limit
based on relation (\ref{eq:Lxyrel}),
\begin{equation}
\label{eq:lxycond}
\delta(L_x(\mbox{A})+L_x(\mbox{B}))^2+
\delta(L_y(\mbox{A})+L_y(\mbox{B}))^2 \geq \frac{7}{8}.
\end{equation}
This inequality requires only two measurement settings
corresponding to 18 measurement probabilities for its
verification. For comparison, the experimental verification
of a Bell's inequality violation reported in \cite{How02}
required four settings and 36 measurement probabilities. 
Moreover, the optimization of the Bell's inequality violations
required measurements at additional settings, while the
measurement settings for the local uncertainty relation
(\ref{eq:lxycond}) are defined by the symmetry of the 
experimental setup and do not have to be varied.
Unfortunately, the measurement data given in \cite{How02}
is not sufficient to allow an analysis of the local
uncertainties of this entanglement source. However, the 
measurement result was interpreted using a simplified 
noise model given in the $\hat{L}_x$ basis by
\begin{eqnarray}
\label{eq:noisemodel}
\hat{\rho}&=& p_s(\mid \mbox{singlet} \rangle \langle
\mbox{singlet}\mid) 
\nonumber \\ && 
+ \frac{(1-p_s)}{3}\left(
\mid +1;-1\rangle \langle +1;-1 \mid
+\mid 0;0\rangle \langle 0;0 \mid
+\mid -1;+1\rangle \langle -1;+1 \mid
\right),
\end{eqnarray}
that is, the correlation along the x-axis of the spin is assumed
to be perfect, while the other two correlations fluctuate with
\begin{eqnarray}
\delta(L_x(\mbox{A})+L_x(\mbox{B}))^2 &=& 0
\nonumber \\
\delta(L_y(\mbox{A})+L_y(\mbox{B}))^2 &=& \frac{4}{3} (1-p_s)
\nonumber \\
\delta(L_y(\mbox{A})+L_y(\mbox{B}))^2 &=& \frac{4}{3} (1-p_s).
\end{eqnarray}
For this model, the relative violation of the local uncertainty
relation (\ref{eq:lxycond}) is equal to
\begin{equation}  
C_{\mbox{\small LUR}} = \frac{32 p_s - 11}{21}.
\end{equation}
Using the value of $p_s=0.69$ reported in \cite{How02}, the 
relative violation of relation (\ref{eq:lxycond}) achieved
in this experiment should be equal to 
$C_{\mbox{\small LUR}} = 0.53$. It might be interesting to 
compare this value with direct measurements of local uncertainty 
violations in future experiments. 

\section{Uncertainty violation and concurrence in 2x2 systems}
For two level systems, the uncertainty relations 
(\ref{eq:sigrel}) and (\ref{eq:xyrel}) define two different
criteria for entanglement verification. 
The local uncertainty relation based on (\ref{eq:sigrel}) 
reads
\begin{equation}
\label{eq:sigcond}
\delta(\sigma_1(\mbox{A})+\sigma_1(\mbox{B}))^2+
\delta(\sigma_2(\mbox{A})+\sigma_2(\mbox{B}))^2+
\delta(\sigma_3(\mbox{A})+\sigma_3(\mbox{B}))^2 \geq 4.
\end{equation}
This uncertainty relation is useful in order to identify
the level of singlet state entanglement in a noisy mixture.
It includes all three Pauli matrices and is therefore not
sensitive to any anisotropy in the noise distribution.
The local uncertainty relation based on (\ref{eq:xyrel}) 
reads
\begin{equation}
\label{eq:xycond}
\delta(\sigma_1(\mbox{A})+\sigma_1(\mbox{B}))^2+
\delta(\sigma_2(\mbox{A})+\sigma_2(\mbox{B}))^2 \geq 2.
\end{equation}
This local uncertainty relation can be tested
with only two measurement settings. It may therefore be
useful in cases where it is necessary to test for entanglement
with only a limited number of measurements. Since one of the
three Pauli matrices is not considered, this condition for
separability is sensitive to noise anisotropies. In particular,
it corresponds to (\ref{eq:sigcond}) if the uncertainty
in $\hat{\sigma}_3(\mbox{A})+\hat{\sigma}_3(\mbox{B})$ is close to two, and is more difficult to violate otherwise.

While a precise characterization of experimentally generated 
quantum states is very difficult, a measurement of the 
uncertainties can provide a comparatively simple test of 
an essential entanglement property. A complete illustration 
of the many kinds of errors in entanglement generation that
can increase the uncertainty levels and thus degrade the 
entanglement is beyond the scope of this paper. In fact, the 
uncertainty limits presented above are useful precisely
because they do not require a full characterization of 
the statistics given by the complete density matrix. 
Nevertheless it may be useful to look at one specific 
example to illustrate the relationship between the uncertainty
boundaries and the actual entanglement of the density
matrix. The most simple case is given by a mixture of a 
maximally noisy state and the intended
maximally entangled state often referred to as a Werner state 
\cite{Wer89},
\begin{equation}
\label{eq:werner}
\hat{\rho}= (1-p_s) \frac{1}{4}\hat{1} + 
p_s \mid \mbox{singlet} \rangle \langle \mbox{singlet} \mid.
\end{equation}
Here, the parameter $p_s$ represents the fraction by which the
intended entangled state exceeds the background noise.
For pairs of two level systems, the amount of entanglement of 
any quantum state can be expressed in terms of the concurrence 
$C$ \cite{Hil97,Woo98}. For Werner states, the concurrence is
$C=\mbox{max}\{(3 p_s-1)/2,0\}$.
It is interesting to compare this precise measure of the total
entanglement of the two systems with the relative violations of 
local uncertainty defined by relation (\ref{eq:sigcond}). 
Since the Werner state is completely isotropic, the 
uncertainties of each component 
$\hat{\sigma}_i(\mbox{A})+\hat{\sigma}_i(\mbox{B})$ are given by
\begin{equation}
\delta (\sigma_i(\mbox{A})+\sigma_i(\mbox{B}))^2 = 2 (1-p_s).
\end{equation}
Therefore, relation (\ref{eq:sigcond}), which gives equal weight 
to each component, appears to be optimally suited as a measure
of entanglement for this class of states. This expectation is 
indeed confirmed by the relative violation of local uncertainty, 
which is in this case precisely equal to the concurrence,
\begin{equation}
\label{eq:concurrence}
C_{\mbox{\small LUR}} = 1-\frac{1-p_s}{3} = C.
\end{equation}
This result shows that for some class of states, the 
concurrence is exactly equal to the amount of noise suppression
achieved in the total spin variables. 
It is an interesting question how large this class of states
is. At present, we would like to note that it is straightforward
to extend the result to arbitrary mixtures of Bell states. 
In general, it seems to be quite significant that the relative 
violation of uncertainty can be used as an estimate of the 
concurrence, even though the experimental effort involved in 
any precise determination of the concurrence greatly exceeds 
the effort required to measure the relative violation of local
uncertainty.

In this context, it may also be interesting to consider  
uncertainty relation (\ref{eq:xycond}), which requires only
two measurement settings. Clearly, this uncertainty limit
is more difficult to overcome because it does not include 
the correlations in the third component 
$\hat{\sigma}_3(\mbox{A})+\hat{\sigma}_3(\mbox{B})$. 
As a result, the relative violation of this uncertainty 
for Werner states is lower than the concurrence $C$ by
\begin{equation}
\label{eq:cprime}
C^\prime_{\mbox{\small LUR}} = 1 - \frac{1-p_s}{2} = C - \frac{1-C}{2}.
\end{equation}
However, since the relative violation of (\ref{eq:xycond})
is always lower than the relative violation of 
(\ref{eq:sigcond}), $C^\prime_{\mbox{\small LUR}}$ may provide a 
useful lower bound for an experimental estimate of 
the concurrence using only two measurement settings.

\section{Further possibilities and open questions}

As explained in section \ref{sec:ur},
uncertainty relations can be formulated for any
operator set $\{\hat{A}_i\}$. 
It is therefore possible to optimize the choice of 
operators in the local uncertainty relation with respect
to a given physical situation. In particular, it may
be possible to classify entangled states according to the
types of local uncertainty relations they violate. 
In any case, it should be kept in mind that the examples 
given here are far from complete.

As mentioned in section \ref{sec:lur}, local
uncertainty relations can also be formulated for NxM 
entanglement, where the dimensionality of the two 
Hilbert spaces is different. One application of
this possibility may be the investigation of multipartite
entanglement, where it allows the formulation of 
bipartite uncertainty limits for various partitions
of the multipartite system \cite{foot1}.

As noted in the introduction, local uncertainties may
also be useful as a characterization of the increased 
precision obtained from entanglement in applications such as 
teleportation, lithography, and clock synchronization
\cite{Tsu02,Hof02b,Bot00,Chu00,Joz00}. On the other
hand, quantum information protocols usually define 
entanglement with respect to distillability by local operations
and classical communication. This raises the
question how the two concepts are related to each other.
Does the distillation of entanglement actually decrease the
uncertainty in the non-local correlation, or does it merely
redistribute the quantum fluctuations \cite{foot2}? 

These are just a few of the questions raised by the 
possibility of quantifying the violation of local uncertainty 
relations by entangled states. A systematic classification of 
local uncertainty relations may thus provide many new insights
into the physical properties of entangled states.

%
\section{Conclusions}
In conclusion, we have generalized the uncertainty
principle to uncertainty sums of arbitrary sets of physical
properties and derived local uncertainty relations
valid for all separable states of a pair of N-level quantum 
systems. Any violation of these local uncertainty relations 
indicates that the two systems are entangled.
The relative violation of a local uncertainty provides
a quantitative measure of this entanglement property and
may be used to evaluate experimental entanglement generation 
processes. It should also be possible to obtain valid estimates
of the total entanglement from uncertainty measurements. 
Specifically, the relative violation of local uncertainty is
actually equal to the concurrence for some 2x2 cases.
In more general cases, it may be possible to identify the
minimal amount of entanglement necessary to obtain the
observed level of local uncertainty violation, thus 
establishing a more precise relation between the local
uncertainty violation and the total entanglement of the
system. Local uncertainty relations may thus provide an
interesting starting point for further investigations into
the physical properties of entanglement.

\end{document}